\documentclass[showpacs,preprint,amssymb]{revtex4}
\usepackage{graphicx}% Include Figure files
\usepackage{dcolumn}% Align table columns on decimal point
\usepackage{bm}% bold math
\def\be{\begin{equation}} 
\def\ee{\end{equation}}
\def\bea{\begin{eqnarray}} 
\def\eea{\end{eqnarray}}
\def\line{\hbox to \hsize}    
\def\frac #1#2{{#1\over #2}}

\def\tr{{\rm  tr\,}}

\def \x{{\bf x}}

\def \a{{\bf a}}

\def \ket #1{{\vert #1\rangle}}

\def\eval #1#2#3{{\langle#1\vert#2\vert#3\rangle}} 

\def\1{\mbox{\bf 1}}
\def\bm#1{\mbox{\boldmath$#1$}} 

\begin{document}
%\draft %(only for revtex) 

\title{A Classical Version of the Non-Abelian Gauge Anomaly}

\author{ MICHAEL STONE}

\affiliation{University of Illinois, Department of Physics\\ 1110 W. Green St.\\
Urbana, IL 61801 USA\\E-mail: m-stone5@illinois.edu}   

\author{VATSAL DWIVEDI}

\affiliation{University of Illinois, Department of Physics\\ 1110 W. Green St.\\
Urbana, IL 61801 USA\\E-mail: vdwived2@illinois.edu}

\begin{abstract}  

We show that a version of the  covariant  gauge anomaly for a 3+1 dimensional chiral fermion interacting with a non-Abelian gauge field can be obtained from the classical  Hamiltonian flow of  its probability distribution in phase space. The only quantum input needed is the Berry phase that arises from the direction of  the spin being slaved to the particle's momentum.

\end{abstract}


\maketitle

\section{Introduction}

There has been much recent interest on the influence of Berry phases on the electronic property of solids \cite{niuRMP}, and a number of these effects provide  fruitful analogies for  relativistic field theories.  
A  particular example occurs when  there is a net flux of Berry curvature through a disconnected part of the  Fermi surface. In this case an analogue of the  Abelian axial anomaly appears, manifesting  itself   as non-conservation of conduction-band particle number in the presence of  external electric and magnetic fields \cite{son1,son2,chen-pu-wang-wang}.    A net Berry flux through the Fermi-surface implies the existence of a  Dirac-cone  band-touching point somewhere within the surface, and the  lower  (valence) band is the  source of the new particles.   In a bulk crystal  the Nielsen-Ninomiya  theorem \cite{nielsen-ninomiya}   requires   that  Dirac-cone  degeneracies come in pairs with opposite-sign  anomalies. Consequently, while the number of particles in each disconnected  Fermi sea will change,   the total number of particles in the conduction and valance band is conserved. On the surface of a topological insulator, however,    we can  have domain-wall fermions \cite{boyanovsky,kaplan} with single Dirac points, and in that case  the additional particles flow into the surface-state  valance band from the bulk  {\it via\/}  the Callan-Harvey effect \cite{callan-harvey}.   

The axial anomaly  is usually   derived  {\it via\/} sophisticated  quantum calculations, so  it is perhaps  surprising that Stephanov and Yin  were able to obtain  the result of  \cite{son1,son2,chen-pu-wang-wang}   from  purely  classical Hamiltonian  phase-space dynamics    \cite {stephanov}.   Their  argument works  because  near the  Fermi surface,   and well  away from the Dirac point, an adiabatic  classical approximation becomes sufficiently accurate  that   the influx of extra particles to  the Fermi surface can be  counted  reliably. The only quantum input is the Berry phase, which subtly alters the classical canonical structure so that ${\bf p}$ and ${\bf x}$ are no longer conjugate variables.

In addition to the Abelian axial anomaly, chiral fermions may also be subject to   a { non-Abelian} {\it gauge\/} anomaly,  in which the failure of a   current to be { covariantly\/}  conserved    signals a quantum breakdown of  the formal gauge invariance.  Since  a  {\it covariant\/}  conservation law  does not imply that any net charge is time independent,  its failure is not necessarily due to   an influx of discrete particles. Instead,  it  means that  the fermion determinant is no longer a function but has become   a  section of a twisted line bundle over the space of gauge-equivalent fields \cite{gaume}.  It might  not  be expected, therefore,  that the gauge anomaly  can be accounted for as  simply as the Abelian axial anomaly. This raises the question of   what the   analogous    phase-space calculation   reveals for  particles  coupled to non-Abelian gauge fields.    The purpose of this paper is to show that  classical phase space dynamics  does  in fact lead to   a   version of the gauge anomaly.

In section \ref{SEC:abelian} we provide a brief  review of the argument in \cite{stephanov}. We do so  because the Abelian calculation  provides a guide  for the slightly more intricate non-Abelian dynamics.  In section     \ref{SEC:kirillov} we review the  classical-quantum correspondence for Lie group representations. In section  \ref{SEC:non-abelian} we derive Liouville's theorem  for the Hamiltonian flow in the  combined gauge and space-time phase space, and show how it leads to a classical analogue of the non-Abelian gauge anomaly. A last section provides a brief  discussion. 

\section{Liouville's Theorem and the  Abelian Anomaly}
\label{SEC:abelian}

In \cite{stephanov} the authors show that the adiabatic  motion of a    3+1 dimensional  positive-energy,  positive-helicity Weyl particle  may be  described by the action functional  
\be
S[{\bf x},{\bf p}] = \int  dt\left({\bf A}\cdot \dot{\bf x} - \phi({\bf x}) +{\bf p}\cdot {\dot {\bf x}}-|{\bf p}| -\a\cdot {\dot {\bf p}}\right).
\label{EQ:Abelian-action}
\ee
Here ${\bf A}$ and $\phi$ are the usual Maxwell vector and scalar potentials.
The vector  potential  ${\bf a}({\bf p})$ is  the real-valued  momentum-space Berry connection that arises because the  Weyl Hamiltonian
\be
H_{\rm Weyl}={\bm \sigma}\cdot {\bf p}
\ee
slaves   the  the spin   of an energy $E=+|{\bf p}|$ particle to  the direction of its  momentum.  The Berry connection is singular:
\be 
{\bf b}=\nabla_{\bf p} \times {\bf a}= \frac{\hat {\bf p}}{2|{\bf p}|^2},\quad \nabla \cdot {\bf b}=2\pi \delta^3({\bf p}).
\ee
The Dirac point ${\bf p}={\bf 0}$ is a Berry-curvature Dirac monopole.

An  adiabatic approximation  has been made in (\ref{EQ:Abelian-action}) that subsumes  all the of the effects of the particle's spin into the Berry phase.   This  approximation  breaks down  completely in the neighbourhood of  the Dirac point,  but   becomes better and better as we move to higher energies.  In the following calculations the Dirac point may   seem to be exactly where  the anomaly calculation needs  the approximation  to hold. This is not the case  however.   As explained in \cite{stephanov}, all that  is important is that the approximation and its  resulting classical flow equations be reliable at a  positive-energy  Fermi surface.

From the action we obtain the classical equations of motion.
\bea
\dot {\bf p}&=& {\bf E} +\dot {\bf x}\times {\bf B}\nonumber\\
\dot {\bf x}&=&\hat {\bf p} +\dot {\bf p}\times {\bf b}
\label{EQ:AbelianEOM}
\eea
The first  equation is the usual Lorentz force. The second  contains the expected    $\hat {\bf p}=\nabla_{\bf p}|{\bf p}|$ group velocity,   but in addition  there is an   anomalous velocity term $\dot {\bf p}\times {\bf b}$. This  term was first identified   by Karplus and Luttinger \cite{luttinger1} who  argued  that it  was   responsible for the anomalous Hall effect in ferromagnetic solids  \cite{blount}.   They were writing thirty  years before the wide-ranging importance  of the adiabatic phase  was made clear by Berry, and   their claim   was not understood and  accepted   until relatively recently   \cite{sundaram-niu,ong,haldane,nagaosa}.

The equations (\ref{EQ:AbelianEOM}) can be solved for $\dot {\bf x}$, $\dot {\bf p}$ in terms of ${\bf x}$ and ${\bf p}$,  to give.
\bea
(1+{\bf b}\cdot {\bf B})\dot {\bf x}&=& \hat {\bf p} + {\bf E}\times {\bf b} + ({\bf b}\cdot \hat{\bf p}) {\bf B},\nonumber\\
(1+{\bf b}\cdot {\bf B})\dot {\bf p}&=& {\bf E} + \hat {\bf p}\times {\bf B} + ({\bf E}\cdot {\bf B}) {\bf b}.
\eea
The phase space $(\dot {\bf x},\dot {\bf p})$ flow is  Hamiltonian, albeit with an unconventional symplectic structure \cite{duval} in which ${\bf p}$ is no longer the canonical conjugate of ${\bf x}$.  We can therefore find a version of  Liouville's theorem for  the conservation of phase-space volume.

We set 
$\sqrt{G}= 1+{\bf b}\cdot {\bf B}$, and   use the  homogeneous Maxwell equations $\nabla_{\bf x} \cdot {\bf B}=0$ and $\nabla_{\bf x}\times {\bf E}+\dot{\bf  B}=0$, to evaluate 
\bea
&&\!\!\!\!\frac{\partial \sqrt{G}}{\partial t}+\frac{\partial \sqrt{G} \dot x_i}{\partial x_i} + \frac{\partial {\sqrt G}\dot p_i}{\partial p_i}\nonumber\\
&=&
{\bf b}\cdot \dot {\bf B}
+ {\bf b} \cdot (\nabla \times {\bf E})+ ({\bf b}\cdot \hat{\bf p})\nabla_{\bf x} \cdot {\bf B}+
({\bf E}\cdot {\bf B})\nabla_{\bf p}\cdot {\bf b}.\nonumber\\
&=&
({\bf E}\cdot {\bf B})\nabla_{\bf p}\cdot {\bf b}.
\label{EQ:Abelian-liouvillle}
\eea
For a non-singular Berry connection  $\nabla_{\bf p}\cdot {\bf b}=0$, and (\ref{EQ:Abelian-liouvillle}) shows that the  conserved phase-space measure is
\be
\mu= \sqrt{G} \left(\frac{dpdx}{2\pi}\right)^3.
\ee
This  measure with its    $\sqrt{G}$ modification was originally regarded as ``non-canonical''   \cite{niu}, but  it is precisely the  canonical   phase-space volume  associated with the unconventional symplectic structure \cite{duval-prl}.  

If we introduce  a phase space density  $f({\bf x},{\bf p},t)$ and  define 
\bea
\rho &=& (1+({\bf b}\cdot {\bf B}))f\nonumber\\
{\bf j}_{\bf x}&=&(1+({\bf b}\cdot {\bf B}))f \, \dot {\bf x}\nonumber\\
{\bf j}_{\bf p}&=& (1+({\bf b}\cdot {\bf B}))f\, \dot {\bf p},
\eea
then, again in the non-singular case, we have 
\bea
&&\frac{\partial \rho}{\partial t}+ \nabla_{\bf x}\cdot {\bf j}_{\bf x}+\nabla_{\bf p}\cdot {\bf j}_{\bf p}\nonumber\\
&=&(1+({\bf b}\cdot {\bf B}))\left( \frac{\partial }{\partial t}+\dot  {\bf x}\cdot \nabla_{\bf x}+ \dot  {\bf p}\cdot \nabla_{\bf p}\right)f.
\label{EQ:Abelian-advect}
\eea
This shows  that  the phase-space probablity density $\rho=\sqrt{G}f$ is conserved when $f$ is advected with the flow. As a consequence of the probability conservation  the particle number $4$-current 
\bea
J^0({\bf x},t) &=&\int f({\bf x},{\bf p},t)\sqrt{G} \frac{d^3p}{(2\pi)^3},\nonumber\\
J^i({\bf x},t) &=& \int f({\bf x},{\bf p},t)\dot x^i \sqrt{G} \frac{d^3p}{(2\pi)^3},
\eea
is   also conserved.
In the singular case, with its  Dirac monopole,  we instead find  
\be
\partial_\mu J^\mu= \frac{1}{(2\pi)^2} ({\bf E}\cdot {\bf B}) f({\bf x},{\bf 0},t).
\label{EQ:Abelian-anomaly}
\ee
If the negative-energy Dirac sea is completely  filled, and in addition some of the positive energy states are filled up a Fermi energy, then we will have  $f({\bf x},{\bf 0},t)=1$, and   equation (\ref{EQ:Abelian-anomaly}) becomes  the 3+1 dimensional  axial anomaly for  positive-chirality  particles.   (For a negative chirality  particle,  the Berry phase and anomaly have opposite sign.)

\section{Classical Mechanics of  Group Representations}
\label{SEC:kirillov}

We will   generalize the  Abelian calculation by making use of the  Wong equations \cite{wong} for a  particle interacting with a Yang-Mills field.  This requires us to appreciate  that  
the   ``charge''  of a particle interacting  with a non-Abelian gauge field is    the  representation $\Lambda$ of the gauge group $G$ in which the particle lives. To  obtain a classical version of the particle's motion in physical space,  the  internal colour space  must also  be  described classically. To do this  the   finite-dimensional   representation  space $\Lambda$ should  be  replaced by a suitable finite-volume phase space ${\mathcal O}_{\Lambda}$  \cite{balachandran}.    

The correspondence  between  Lie group representations, classical phase space and quantization has been explored in great generality  by Kirillov, Kostant and Souriau  \cite{kirillov}. We   will, however,  restrict ourselves  to the specific    case of  a compact simple group. For such a group a unitary irreducible representation is completely characterized by its highest-weight vector $\ket{\Lambda}$. The  general theory in \cite{kirillov} then shows that the appropriate  
phase space ${\mathcal O}_\Lambda$ is   the co-adjoint  orbit of the function  $F(X)= \eval{\Lambda}{X}{\Lambda}$  under the map $F(X)\to F(g^{-1}Xg)$.  Here $g^{-1}Xg$  denotes the adjoint action of $g^{-1}$ on the Lie algebra element $X$. Now a  simple  group possesses an invertible Killing-form metric tensor that we can take to be 
\be
\gamma_{ab}=\tr\{\lambda_a\lambda_b\},
\ee
where the trace is taken is some fixed faithful representation (usually the defining representation) of $G$, and the $\lambda_a$ are a hermitian basis for the Lie algebra obeying $[\lambda_a,\lambda_b]=i{f_{ab}}^c \lambda_c$. Making use of this  metric allows us to write 
\be
 \eval{\Lambda}{g^{-1}Xg}{\Lambda}=\tr\{\alpha_\Lambda  g^{-1}Xg\}= \tr\{g\alpha_\Lambda  g^{-1}X\},
 \label{EQ:orbits}
\ee
where  
\be
\alpha_\Lambda=\alpha^a_\Lambda \lambda_a, \qquad \alpha^a_\Lambda =\gamma^{ab} \eval{\Lambda}{\hat \lambda_b}{\Lambda}.
\label{EQ:weight}
\ee
Here $\hat\lambda_a$ is the matrix representing the generator $\lambda_a$ in the representation $\Lambda$.  In the compact simple case therefore, the second equality in (\ref{EQ:orbits}) shows that the co-adjoint orbit of $F$ can be identified with the adjoint orbit of $\alpha_\Lambda$.

Consider now the  Hamiltonian action functional
\be
S[ g]= \int dt \left( i\,\tr\left\{ \alpha g^{-1}\frac{dg}{dt}\right\}- i\,{\mathcal H}(g)\right),
\ee
where $\mathcal{H} (g)= \tr\{\alpha g^{-1}Xg\}$ with $X$   an element of the  Lie algebra.
The equation of motion that comes  from varying $g$   is 
\be
[\alpha, g^{-1}(\partial_t -X)g]=0.
\ee
This equation  is equivalent to 
\be
g^{-1}(\partial_t -X)g +h(t)=0,
\ee
where $h(t)$ is an arbitrary time dependent function such that $h\in {\mathfrak g}\equiv {\rm Lie}(G)$ commutes with $\alpha$.
The solution to the equation of motion is  therefore     
\be
g(t) = {\mathcal T}\exp\left\{ \int^t _0 X dt\right\}H(t),
\ee
where $\dot H(t)=h(t)$ is now  an arbitrary element of the subgroup $H\subseteq G$ that commutes with $\alpha$.  The  group element $g(t)$ is  thus  only well defined as an element of the coset  $G/H$.
The Lie-algebra valued expression 
\be
Q= g\alpha g^{-1}=Q^a\lambda_a
\ee
is insensitive to the $H(t)$ ambiguity, and its  (co)-adjoint orbit  can be identified with the coset  $G/H$. This coset is a our phase space ${\mathcal O}_\Lambda$, and  we can  define a Poisson bracket on functions on ${\mathcal O}_\Lambda=G/H$ by setting
\be
\{{\mathcal H}_1,{\mathcal H}_2\}\stackrel{\rm def}{=}\left. \frac{d {\mathcal H}_2}{dt}\right|_{{\mathcal H}_1} .
\ee
In particular,    $Q_a= \gamma_{ab}Q^b= \tr\{Q\lambda_a\}= \tr\{\alpha g^{-1} \lambda_a g\}$ is a function on $G/H$,  and   we find that 
\be
\{Q_a,Q_b\}=i{f_{ab}}^c Q_c.
\ee
This Poisson-bracket version of the Lie algebra  exists  for any $\alpha\in{\mathfrak g}$,  but only when  $\alpha$ arises as an   $\alpha_\Lambda$ from equation (\ref{EQ:weight})  can the classical motion be consistently quantized. When  we do so, we  recover the representation $\Lambda $. In this case 
the classical-variable $\to$ quantum-operator  correspondence will assign   $Q_a \to  \hat \lambda_a$, where $\hat \lambda_a$ is the matrix  representing $\lambda_a$ in the representation $\Lambda$.    

In addition to  hosting  a classical version of the Lie algebra, the coadjoint orbit   provides a classical version of the   the symmetrized trace of the quantum operators $\hat \lambda_a$  as 
\be
{\rm str}_{\Lambda}(\hat \lambda_{a_1}\cdots \hat \lambda_{a_n})\sim \int_{{\mathcal O}_\lambda} Q_{a_1}\cdots Q_{a_n} \,\mu_\Lambda,
 \ee
where
\be
\mu_\Lambda =\frac 1{N!(2\pi)^N} (-\tr\{\alpha_\Lambda (\omega_L)^2\})^N 
\ee
is the canonical measure on the phase space. The  symbol 
 $\omega_L=\omega_L^a \lambda_a =g^{-1}dg$  denotes the left-invariant Maurer-Cartan form on the Lie algebra, and $N$ is the number of pairs of generators that fail to commute with $\alpha$.   
 
The simplest such trace  integral would be   
\be
\int_{{\mathcal O}_\Lambda} 1\,  \mu_{\Lambda} \stackrel{?}{=} \tr_\Lambda \{\mathbb I\}={\rm dim}(\Lambda).
\label{EQ:normalize}
\ee
However it is known  \cite{kirillov} that  to get the {\it exact\/}   dimension  we must  integrate not over   ${\mathcal O}_\Lambda$ but over the slightly larger orbit  ${\mathcal O}_{\lambda+\rho}$ where $\rho$ is the Weyl vector (half the sum of the positive roots of the algebra, or equivalently the sum of the fundamental weights).  
 For $G={\rm SU}(2)$, for example,  we can take the generators in the fundamental representation to be $\sigma_i$ and then  the spin-$j$ representation has $\alpha_j =j\sigma_3$ with  $N=1$. Using  Euler angles to parameterize the group element 
\be
g= \exp\{-i\phi\sigma_3/2\}\exp\{-i\theta\sigma_2/2\}\exp\{-i\psi\sigma_3/2\},
\ee
and setting  $g^{-1}dg = \omega_L^a\sigma_a$, we have 
\bea
 \omega_{\rm L}^1 &=& -\frac{i}{2}(\sin\psi\, d\theta -\sin\theta \cos\psi\, d\phi),\nonumber\\
\omega_{\rm L}^2 &=&-\frac{i}{2}( \cos\psi\, d\theta +\sin\theta \sin\psi\, d\phi),\nonumber\\
\omega_{\rm L}^3 &=&-\frac{i}{2}( d\psi +\cos\theta\, d\phi).
\eea
This gives 
\bea
-j\tr\{\sigma_3 (\omega_L)^2\}/2\pi &=&-2j(\omega_L^1\omega_L^2 -\omega_L^2\omega_L^1)/2\pi \nonumber\\
&=& - 4j\,\omega^1_L\omega_L^2/2\pi \nonumber\\
&=& \frac{j}{2\pi}\sin\theta \,d\theta \,d\phi.
\eea
The integral over the 2-sphere  orbit therefore gives
\be
\int_{{\mathcal O}_j} \mu_j=2j,
\ee
which is not quite right.
The Weyl shift  replaces $j$ with $ j+1/2$ and so yields  the correct dimension of $2j+1$.   In the appendix we  consider  ${\rm SU}(3)$ and show  that 
\be
\int_{{\mathcal O}_{\Lambda+\rho}} Q_{a}Q_{b}Q_{c}\,\mu_{{\Lambda+\rho}} = \frac 12 \tr_{\Lambda}(\lambda_a\{\lambda_b,\lambda_c\})
\label{EQ:three-trace}
\ee
for all representations. The shift by $\rho$ is a quantum correction that can have a large effect for smaller representations. It becomes less significant for larger and more classical representations.

\section{Liouville's Theorem and the   Gauge Anomaly}
\label{SEC:non-abelian}
 
We  now couple the internal group dynamics  to the motion  of our   3+1 dimensional  Weyl fermion.    We take as action functional 
\be
S[{\bf x},{\bf p}, g]= \int dt\left(i\, \tr\left\{ \alpha g^{-1}\left(\frac{d}{dt} -i(\dot \x \cdot {\bf A}+A_0)\right)g\right\}+{\bf p}\cdot {\dot {\bf x}}-|{\bf p}| -\a\cdot {\dot {\bf p}}\right).
\label{EQ:wong-action}
\ee
Here  the  factors of $i$  have been inserted so that the non-Abelian gauge fields $A_0=A_0^a\lambda_a$, ${\bf A}={\bf A}^a\lambda_a$ are hermitian. Recall  that   for  Abelian electromagnetism we have $A_0 =-\phi$.  In the Abelian  case, therefore,  the action reduces to (\ref{EQ:Abelian-action}).
The functional  (\ref{EQ:wong-action}) is invariant under the gauge transformation
\bea
-iA_\mu&\to& -iA_\mu ^h= h^{-1}(-iA_\mu)h +h^{-1}\partial_\mu h,\nonumber\\
g(t)&\to& h^{-1}(\x(t),t)g(t).
\eea

From (\ref{EQ:wong-action}) we obtain the  equation of motion for $g$:  
\be
[\alpha, g^{-1}(\partial_t -i\dot \x \cdot {\bf A}-i A_0)g]=0.
\ee
As before, $g(t)$ is only defined as an element of $G/H$.
For constant $C$, however,  we have the unambiguous result
\bea
\frac{\partial}{\partial t} \tr\{QC\}&=&\tr\{[Q,-iA_0-i\dot \x\cdot {\bf A}]C\}\nonumber\\
&=& \tr\{Q[-iA_0-i\dot \x \cdot {\bf A},C]\}
\eea
We use this  result when we vary ${\bf x}$ to get the equation of motion
\bea
\dot p_i &=& -\frac{\partial}{\partial t} \tr\{QA_i\}+ \dot x_j \frac{\partial}{\partial x_i}\tr\{QA_j\} + \frac{\partial}{\partial x_i} \tr \{Q A_0\}\nonumber\\
&=& \tr\left\{Q\left(\left(\frac{\partial A_0}{\partial x_i}-\frac{\partial A_i}{\partial t} -i[A_i,A_0]\right)+ \dot x_j\left(\frac{\partial A_j}{\partial x_i}-\frac{\partial A_i}{\partial x_j}-i[A_i,A_j]\right)\right)\right\}.
\eea
We have  obtained  the equation derived empirically by Wong \cite{wong}
\be
\dot {\bf p} =\tr\{Q ({\bf E}+\dot {\bf x}\times {\bf B})\}.
\ee
Here ${\bf E}={\bf E}^a\lambda_a$ and ${\bf B}={\bf B}^a\lambda_a$ are the Lie-algebra-valued  non-Abelian analogues of the electric and magnetic fields whose  $i=1,2,3$  components are 
\bea
B_i&=& \frac 12 \epsilon_{ijk}\{\partial_j A_k-\partial_kA_j -i[A_j,A_k]\},\nonumber\\
E_i&=& \partial_i A_0- \partial_0A_i-i[A_i,A_0].
\eea 

By  varying  ${\bf p}$ we again get 
\be
\dot \x=  \hat {\bf p} +\dot {\bf p}\times {\bf b}.
\ee
The full set of equations determining the motion   is  therefore
\bea
\dot Q&=& -i[Q, A_0+\dot \x\cdot {\bf A}],\nonumber\\
\dot {\bf p} &=&\tr\{Q ({\bf E}+\dot \x\times {\bf B})\},\nonumber\\
\dot {\bf x}&=&  \hat {\bf p} +\dot {\bf p}\times {\bf b}.
\eea
These may again be solved for  $\dot {\bf x}$ and $\dot {\bf p}$ in terms of ${\bf x}$, ${\bf p}$ and $Q$, as 
\bea
(1+{\bf b}\cdot \tr\{Q {\bf B}\})\dot {\bf x}&=& \hat {\bf p} + \tr\{Q{\bf E}\}\times {\bf b} + ({\bf b}\cdot \hat{\bf p})
 \tr\{Q{\bf B}\}.\nonumber\\
(1+{\bf b}\cdot \tr\{Q {\bf B}\})\dot {\bf p}&=& \tr\{Q{\bf E}\} + \hat {\bf p}\times \tr\{Q{\bf B}\} + (\tr\{Q{\bf E}\}\cdot \tr\{Q{\bf B}\}) {\bf b}.
\eea

A reasonable conjecture is that   the non-Abelian generalization of the phase-space measure involves  $\sqrt{G}=(1+{\bf b}\cdot \tr\{Q {\bf B}\})$.  A  slightly tedious computation with the  symplectic form  confirms that this conjecture is correct,  and the measure is 
\be
\mu= (1+{\bf b}\cdot \tr\{Q {\bf B}\}) \mu_\Lambda \left(\frac{dpdx}{2\pi}\right)^3.
\ee

To obtain the non-Abelian version of Liouville's theorem
we  need   the analogues 
\bea
\nabla\cdot {\bf B} -i ({\bf A}\cdot {\bf B}- {\bf B}\cdot {\bf  A})&=&0, \nonumber\\
\dot {\bf B}-i [A_0, {\bf B}]+ \nabla \times {\bf E} -i({\bf A}\times{\bf E}+{\bf E}\times {\bf A})&=&0,
\eea
of the homogeneous Maxwell equations.
These  homogeneous equations lead to   the sum  of the three terms 
\bea
\left(\frac{\partial \sqrt{G}}{\partial t}\right)_{Q} +{f_{ab}}^c Q^a A^b_0 \left(\frac{\partial \sqrt{G}}{\partial Q^c }\right)_{t,{\bf x}\,{\bf p}}  
&=& {\bf b} \cdot \tr\{Q(\dot {\bf B}-i[A_0, {\bf B}])\},\nonumber\\
\left(\frac{\partial \sqrt{G}\dot x^i }{\partial x^i}\right)_{Q} +{f_{ab}}^c Q^a A^b_i \left(\frac{\partial \sqrt{G}\dot x^i}{\partial Q^c }\right)_{t, {\bf x},{\bf p}}  &=& \epsilon_{ijk}\tr\{Q(\partial_i E_j-i[A_i,E_j]\}b_k ,\nonumber\\
&& \quad\qquad +(\hat {\bf p}\cdot {\bf b})\tr\{Q(\partial_i B_i -i[A_i, B_i]\},\nonumber\\
\left(\frac{\partial \sqrt{G}\dot p_i}{\partial p_i} \right)_{Q,t, {\bf x}}&=& \tr\{Q{\bf E}\}\cdot \tr\{Q{\bf B}\}\nabla\cdot {\bf b},
\label{EQ:non-Abelian-liouville1}
\eea
 being   zero   ---   modulo the singular contribution  from   $\nabla \cdot {\bf b}$ in the last line. This is our   non-Abelian  Liouville theorem. The theorem   can also be derived with a bit more effort by  computing   the Lie derivative of the top power of the symplectic form. A tricky point here is the computation of the Lie derivative of $\mu_\Lambda$. This  derivative is not zero, and is responsible  for moving an  $\dot x^i$  in the second line of (\ref{EQ:non-Abelian-liouville1}) away from its natural companion  $A^b_i$ to its location   inside the $Q^c$ derivative.  

To understand  why this rather complicated looking relation is an expression of  phase-space conservation  we observe that 
the convective constancy   of a phase-space  distribution $f(Q^a,{\bf x},{\bf p},t)$ is  expressed by
\be
\left(\frac{\partial}{\partial t}+ \dot x^i \frac{\partial}{\partial x^i}+\dot Q^a\frac{\partial}{\partial Q^a}+\dot p_i \frac{\partial}{\partial p_i}\right)f=0.
\label{EQ:vlasov}
\ee
Now 
\be 
\dot Q^a= -{f_{bc}}^a (A^b_0+\dot x^i A^b_i) Q^c,
\ee
so  we can group the $\dot Q$, with the  $t$ and $x$ derivatives together to make two     $A$-dependent ``covariant derivatives'' \cite{litim}.  These are
\be
\left(\frac{\partial f}{\partial t}\right)_Q +{f_{ab}}^cQ^aA_0^b \left(\frac{\partial f }{\partial Q^c}\right)_{t,{\bf x},{\bf p}}
\ee
and
\be
\dot x^i\left(\left(\frac{\partial f }{\partial x^i}\right)_Q+{f_{ab}}^cQ^a{A}_i^b \left(\frac{\partial f  }{\partial Q^c}\right)_{t,{\bf x},{\bf p}}\right).
\ee
We see that we have  the same combination of terms that we had in (\ref{EQ:non-Abelian-liouville1}).
Let us verify that these combinations  are naturally  gauge covariant.
Under a transformation $Q\to Q'= gQg^{-1}$ we have
\be
Q^a\to {Q'}^a ={G^a}_b Q^b
\ee
where  ${G^a}_b\equiv {[{\rm Ad}(g)]^a}_b$ is the matrix corresponding to $g$ in  the adjoint representation of $G$.  Since $g({\bf x},t)$ depends on space and time, this transformation mixes up the $Q$ and ${\bf x}$ derivatives.  The density  $f$ is  invariant,   
\be
f(Q, {\bf x},{\bf p},t)=f'(Q',{\bf x}, {\bf p},t),
\ee
but   
\be
\left(\frac{\partial f}{\partial t}\right)_Q= \left(\frac{\partial f'}{\partial t}\right)_{Q'}+  \left(\frac{\partial Q'^b}{\partial t}\right)_Q\left(\frac{\partial f'}{\partial Q'^b}\right)_t,
\ee
and 
\be
\left(\frac{\partial f}{\partial Q^a}\right)_t=\left(\frac{\partial Q'^b}{\partial Qa}\right)_t\left(\frac{\partial f'}{\partial Q'^b}\right)_t.
\ee
So we have 
\bea
&&\left(\frac{\partial f}{\partial t}\right)_Q +{f_{bc}}^aA_0^cQ^b \left(\frac{\partial f }{\partial Q^a}\right)_t\nonumber\\
&=& \left(\frac{\partial f'}{\partial t}\right)_{Q'}+ \left\{{(\dot GG^{-1})^b}_d +{G^b}_e {f_{bc}}^e A_0^c {(G^{-1})^c}_d\right\}Q'^d \left(\frac{\partial f'}{\partial Q'^b}\right)_t.
\eea
We  therefore have  covariance  under  
\bea
Q^a&\to& {Q'}^a ={G^a}_b Q^b\nonumber\\
{f_{bc}}^aA_\mu^c&\to&   {f_{bc}}^a A'^c_\mu =   {(\dot GG^{-1})^a}_b +{G^a}_d {f_{ec}}^d A_\mu^c {(G^{-1})^e}_b.
\eea
As   the matrix representing $\lambda_c$ in the adjoint representation is  ${[{\rm ad}(\lambda_c)]^a}_b= -i {f_{bc}}^a$,  this is  indeed  the correct gauge transformation.

Combining Liouville's theorem with (\ref{EQ:vlasov}) shows that $\sqrt{G}f$ is the  conserved (modulo the singular contribution) phase-space probability:   
\be
\frac{\partial \sqrt{G} f}{\partial t} + \frac{\partial \sqrt{G}  f\dot x^i }{\partial x^i} +\frac{\partial \sqrt{G}  f\dot p^i }{\partial p^i}=  f(Q,{\bf x},{\bf p},t)\tr\{Q{\bf E}\}\cdot \tr\{Q{\bf B}\}\nabla\cdot {\bf b}.
\ee

Now we  define the gauge $4$-current
\bea
J^{0}_a(t,{\bf x})&=& \int Q_a f(Q,{\bf x}, {\bf p})\,  \sqrt{G} \mu_\Lambda \frac{d^3p}{(2\pi)^3} ,\nonumber\\ 
J^{i}_a(t,{\bf x}) &=& \int   Q_a \dot x^i f(Q,{\bf x}, {\bf p})\, \sqrt{G} \mu_\Lambda  \frac{d^3p}{(2\pi)^3},
\eea
and combine  Liouville's theorem, the convective constancy of the phase-space density $f(Q^a,{\bf x},{\bf p},t)$,  with  
\be
\dot Q_a= {f_{ba}}^c(A^b_0+\dot x^i A^b_i)Q_c
\ee
to see that
\be
\partial_\mu J^{\mu}_a  - {f_{ba}}^c A_\mu^b J^{\mu}_c =   f(0)  \frac{1}{(2\pi)^2} \int _{{\mathcal O}_\Lambda} Q_a \tr\{Q{\bf B}\}\tr\{Q{\bf E}\} \, \mu_\Lambda. 
\ee
Since $\tr\{Q{\bf B}\}=Q_a {\bf B}^a$, $\tr\{Q{\bf E}\}=Q_a {\bf E}^a$  and the integration of the three factors of $Q_a$ over the phase-space ${\mathcal O}_\Lambda=G/H$ is (up to a  Weyl shift) is the classical version of the symmetrized trace $\frac{1}{2}\tr_\Lambda(\hat\lambda_a\{\hat \lambda_b,\hat \lambda_c\})$, this expression becomes a   classical version  of the ``covariant'' (as opposed to  ``consistent'')  gauge anomaly \cite{bardeen-zumino}
\bea
\nabla_\mu J^{\mu}_a&=& \frac{1}{32\pi^2}\epsilon^{\alpha\beta\gamma\delta} \tr_\Lambda (\hat \lambda_a F_{\alpha\beta}F_{\gamma\delta}),\nonumber\\
&=& \frac{1}{(2\pi)^2} \frac{1}{2} \tr_\Lambda(\hat\lambda_a\{\hat \lambda_b,\hat \lambda_c\}){\bf E}^b \cdot {\bf B}^c.
\label{EQ:covariant-gauge-anomaly}
\eea

\section{Discussion}  

We have  considered  the classical phase space Hamiltonian flow for spin-$\frac 12$ particles interacting with a non-Abelian gauge field. We used  Liouville's theorem to identify  the phase-space volume-form and  found that this  volume-form  fails to be conserved   in the vicinity of the Berry-phase monopole. The failure then leads to a classical version  of the covariant  form of the non-abelian gauge anomaly.  

It is perhaps not too surprising that we  obtain  the ``covariant''   gauge anomaly rather than the ``consistent"  gauge anomaly. Although the Hamiltonian formalism  only makes manifest the    canonical structure,  gauge invariance is being tacitly maintained at all points of the calculation.  Also, when an anomalous chiral gauge theory makes physical sense, the Weyl  particles  will be domain-wall   fermions residing on the boundary of  some higher dimensional space. The anomaly is then accounted for by the inflow of gauge current from the bulk, and this inflowing current can  obtained by functionally differentiating a bulk Chern-Simons action. The boundary variation of the Chern-Simons term is then precisely the Bardeen-Zumino polynomial \cite{bardeen-zumino} that converts the consistent gauge current to the covariant current (see for example    \cite{stone-righi}).  
A similar argument shows that in an anomalous theory   the current that appears in the Lorentz-force contribution to the energy-momentum conservation law is the covariant current \cite{dubovsky,neiman}.

There have been several  recent works on the effects of  anomalies on  fluid dynamics \cite{loganayagam,nair}.  It is interesting to explore the relation between these papers, which take the anomalies  as given and explore their consequences, and the present analysis that uses fluid-like kinetics to deduce their existence.

 \section{Acknowledgements}   This  project was supported by the National Science Foundation  under grant  DMR 09-03291.   MS would like to thank  Peter Horv\'athy for  a valuable  e-mail exchange,  and  Misha Stepanov and Dam Thanh Son for  comments. VD thanks Xiongjie Yu for constructive blackboard discussions.

\appendix
\section{Comparing classical and quantum traces}

 We wish to see how accurately  classical phase-space integrals such as 
\be
\int_{{\mathcal O}_\Lambda} \mu_{\Lambda}, \quad \int_{{\mathcal O}_\Lambda} Q_aQ_b\, \mu_\Lambda  \quad \int_{{\mathcal O}_\Lambda} Q_aQ_bQ_c\, \mu_\Lambda
\ee
(or their Weyl-shifted versions) approximate their respective  symmetrized quantum traces
\be
{\rm dim}(\Lambda)\equiv  \tr_\Lambda(\mathbb I),  \quad \tr_\Lambda(\hat\lambda_a\hat\lambda_b),\quad \frac12 \tr_\Lambda(\lambda_a\{\hat \lambda_b,\hat\lambda_c\}).
\ee

The simplest non-trivial example is provided by  ${\rm SU}(3)$. For ease  in raising and lowering indices we will   normalize the  generators in the fundamental representation so that 
\be
\tr\{\lambda_a\lambda_b\}=\delta_{ab}.
\ee
In particular 
\be
\lambda_3\mapsto  \frac{1}{\sqrt 2} \left(\matrix{1&0&0\cr 0&-1&0\cr 0&0&0}\right),
\ee
and 
\be
\lambda_8\mapsto  \frac{1}{\sqrt 6} \left(\matrix{1&0&0\cr 0&1&0\cr 0&0&-2}\right).
\ee
We now define  the symmetric invariant tensor $d_{abc}$ by 
\be
d_{abc}= \tr(\lambda_a\{\lambda_b,\lambda_c\}). 
\ee
With this definition 
\be
d_{abc}d^{abc}= 2\frac{(n^2-4)(n^2-1)}{n}.
\ee
The factors show  that the $d_{abc}$ coefficients vanish for ${\rm U}(1)$ and ${\rm SU}(2)$.
For ${\rm SU}(3)$ we will only need to know the explicit values 
\be
d_{888}=-\frac{2}{\sqrt6},\quad d_{338}=+\frac{2}{\sqrt 6}.
\ee

%\begin{figure}
%\centerline{
%\includegraphics[width=3.0in]{su3lattice_weyl.eps}
%}
%\caption{\sl The roots and part of the   weight lattice of  $\mathfrak {su}(3)$. The Weyl chamber is highlighted in magneta. }
%\label{FIG:su}
%\end{figure}

The finite-dimensional unitary representations of  ${\rm SU}(3)$ have highest weights (eigenvalues of $\hat \lambda_3$ and $\hat \lambda_8$) that  are non-negative  integer linear combinations $p{\bm \omega}_1+q{\bm \omega}_2$ of the fundamental  weights
\be
{\bm \omega}_1 =\left(\frac{1}{\sqrt 2}, \frac{1}{\sqrt 6}\right),\quad {\bm \omega}_2=\left(0, \frac{2}{\sqrt 6}\right).
\ee
The  representation with highest weight ${\bf \Lambda}= p{\bm \omega}_1+q{\bm \omega}_2$ has dimension
\be
{\rm dim}(p,q)=(p+1)(q+1)(p+q+2)/2.
\ee
From \cite{kirillov} we know that  this dimension is   the volume of ${\mathcal O}_{\Lambda+\rho}$. We can  therefore   read-off that the volume of the unshifted orbit ${\mathcal O}_\Lambda$ is $pq(p+q)$. This unshifted volume goes to zero as we approach  the edges $p=0$ and $q=0$ of the Weyl chamber, where the co-adjoint orbit degenerates and its dimension  reduces from $d=6$ to $d=4$.  The Weyl shift protects us from this degeneration.

With our normalization the quantum quadratic and cubic  Casimir operators
\be
\hat C_2= \hat \lambda_a\hat \lambda_a, \quad \hat C_3=d_{abc}\hat\lambda_a\hat\lambda_b\hat\lambda_c
\ee
 have eigenvalues 
%\footnote{For example:  N.~Arisaka, {\it On the unitary representations of ${\rm SU}(3)$\/}, Prog.\ Theor.\ Phys.\ {\bf 47} 1758-1781 (1972).} \footnote{I am using non-standard definitions. People usually take $T_a=\lambda_a/2$ with $\tr(\lambda_a\lambda_b)=2\delta_{ab}$. They then set $C_2=T_aT_a$ and $C_3=d_{abc}T_aT_bT_c$, where
%$$
%\{T_a,T_b\}= \frac 13 \delta_{ab} {\mathbb I}+d_{abc}T_c.
%$$
%The coefficients in the Casimirs then change $\frac 23\to \frac 13$ and $\frac 29 \to \frac1{18}$.} 
\bea
C_2(p,q)&=&\frac 23 (p^2+pq+q^2+3p+3q),\nonumber\\
C_3(p,q)&=&\frac 29(p-q)(2p+q+3)(2q+p+3).
\eea 
To obtain the  classical  version of these  quantities  we consider  the (co)-adjoint orbit 
\be
Q=g\alpha_\Lambda g^{-1}= Q_a\lambda_a,
\ee
where  the definition (\ref{EQ:weight}) gives
\be
\alpha_\Lambda= \frac{p}{\sqrt 2}\lambda_3 +\left(\frac{p}{\sqrt 6}+\frac {2q}{\sqrt 6}\right)\lambda_8. 
\ee
The invariance of the  tensors $\delta_{ab}$ and $d_{abc}$  ensures that the geometric Casimirs   
\bea
C^{\rm geom}_2&=& \sum_i (Q_i)^2\nonumber\\
C^{\rm geom}_3&=& d_{ijk}Q_iQ_jQ_k
\eea
are constants on the orbit. 
 To compute  these constants we observe that 
the unique point  in the Weyl chamber  at which the co-adjoint orbit intersects the maximal torus  has coordinates
\be
Q^0_3=  \frac{p}{\sqrt 2}, \quad Q^0_8 =\left(\frac{p}{\sqrt 6}+\frac {2q}{\sqrt 6}\right),
\ee 
with all other coordinates vanishing.
The geometric Casimirs, are therefore given by
\bea
C^{\rm geom}_2 =(Q^0_3)^2+(Q^0_8)^2 &=& \frac 23(p^2+pq+q^2)\nonumber\\
C^{\rm geom}_3=d_{888} (Q^0_8)^3+3d_{338} Q^0_8(Q^0_3)^2&=& \frac 29(2p^3+3p^2q-3pq^2-2q^3).\nonumber\\
&=& \frac 29(p-q)(2p+q)(2q+p)
\eea
In other words,  the (co)-adjoint orbit ${\mathcal O}_\Lambda$ is the algebraic curve in ${\mathbb R}^8$   given  by the pair of  equations
\bea
Q_aQ_a&=& \frac 23(p^2+pq+q^2),\nonumber\\
d_{abc}Q_aQ_bQ_c&=& \frac 29(2p^2+3p^2q-3pq^2-2q^3).
\eea
The polynomials in $p$, $q$ in these equations do not   coincide with those giving the quantum Casimirs. Only the highest powers in $p$, $q$ are present.  If, however, we make the Weyl shift ${\bm \Lambda}\to {\bm \lambda}+{\bm \rho}$ ({\it i.e.\/}\ $(p,q)\to (p+1,q+1)$) then the cubic Casmir becomes exact, and the quadratic Casimir is correct up to an additive constant. Indeed, it is easy to show that in  any simple Lie algebra the   geometric  quadratic Casimir is given by 
$C_2^{\rm geom}=|{\bm \Lambda|^2}$  while  the quantum  Casimir is given by  $C_2=|{\bm \Lambda}+{\bm \rho}|^2-|{\bm \rho}|^2$.
The lower powers in the quantum case arise from the necessity of normal-ordering --- {\it i.e.} of commuting all  step-up (positive root) operators to the far right of any expression, where they vanish when acting on  the highest-weight state.

Now, because there is only one rank-two invariant tensor, we   know that
\be
\tr_\Lambda(\hat \lambda_a\hat\lambda_b) = x_\Lambda  \delta_{ab}
\ee
for some number (the integer-valued Dynkin index) $x_\Lambda$. By taking traces we find that
\be
x_\Lambda= \frac{{\rm dim}(\Lambda) C_2}{8}.
\ee
(Here $8=\delta_{aa}= n^2-1$ for $n=3$). 
Similarly, because the measure is invariant under the adjoint action on the orbit, and again because $\delta_{ab}$ is the only invariant two-index tensor we must have 
\be
\int_{{\mathcal O}_{\Lambda+\rho}} Q_aQ_b\, \mu_{\Lambda+\rho}= x_\Lambda^{\rm geom} \delta_{ab}.
\ee
To find  $x_\Lambda^{\rm geom} $ we   contract with $\delta^{ab}$, use $Q_aQ_b\delta^{ab}= C^{\rm geom}_2$, 
and
\be
\int_{{\mathcal O}_{\Lambda+\rho}} \mu_{\Lambda+\rho}= \tr_\Lambda({\mathbb I})={\rm dim}(\Lambda)
\ee
to get
\be
x_\Lambda^{\rm geom}= \frac{{\rm dim}(\Lambda) C^{\rm geom}_2}{8}.
\ee

In the same manner we observe that  there is only one symmetric rank-three invariant tensor, and so  
\be
\frac 12 \tr_\Lambda(\hat \lambda_a\{\hat \lambda_b,\hat \lambda_c\})\,= \frac{{\rm dim}(\Lambda)C_3}{d_{ijk}d^{ijk}}d_{abc}, 
\ee
while 
\be
\int_{{\mathcal O}_{\Lambda+\rho}} Q_aQ_bQ_c\, \mu_{\Lambda+\rho}= \frac{{\rm dim}(\Lambda)C^{\rm geom}_3}{d_{ijk}d^{ijk}}d_{abc}.
\ee
In this case, because $C^{\rm geom}_3= C_3$, the  Weyl-shifted phase-space integral  coincides with the quantum trace for any representation $\Lambda$.

\end{document}